\documentclass[aps,prl,twocolumn,showpacs,superscriptaddress,longbibliography,nofootinbib]{revtex4-2}

\usepackage{amsmath}
\usepackage{amssymb}
\usepackage{graphicx}
\usepackage[colorlinks=true,linkcolor=blue,citecolor=blue, urlcolor=blue]{hyperref}
\usepackage{cleveref}
\usepackage{xcolor}
\usepackage{physics}
\usepackage{xfrac}

\begin{document}

\title{Coalescence as the driver for high $p_T$-hadrons in small systems}
\title{Coalescence versus paths as the driver for high $p_T$-hadrons in small systems}
\title{Radial flow and coalescence explaining high $p_T$-hadrons in small systems}
\title{Radial flow and coalescence explain hadron yield and anisotropy at high $p_T$}
\title{Flow and coalescence explain nuclear modification factor and high-$p_T$ anisotropy}
\title{Flow plus coalescence explain hadron nuclear modification and high-$p_T$ anisotropy}

\author{Wilke van der Schee}
\affiliation{Theoretical Physics Department, CERN, CH-1211 Gen\`{e}ve 23, Switzerland}
\affiliation{Institute for Theoretical Physics, Utrecht University, 3584 CC Utrecht, The Netherlands}
\affiliation{NIKHEF, Science Park 105, 1098 XG Amsterdam, The Netherlands}

\author{Isobel Kolb\'{e}}
\affiliation{School of Physics, University of the Witwatersrand, 1 Jan Smuts Ave, Braamfontein, 2000, South Africa}
\affiliation{Mandelstam Institute for Theoretical Physics, University of the Witwatersrand, 1 Jan Smuts Ave, Braamfontein, 2000, South Africa}
\affiliation{National Institute for Theoretical and Computational Sciences, Merensky building, Merriman Street, Stellenbosch, 7600, South Africa}

\begin{abstract}
It has long been a puzzle why the nuclear modification factor in relativistic $p$Pb
collisions is consistent with unity, while the high-$p_T$ elliptic flow is significantly positive. 
The latter is traditionally interpreted as path-length-dependent energy loss, whereas the former is consistent with zero energy loss. 
In this Letter, we propose that even at high $p_T$ a hard parton coalesces with a boosted thermal medium parton whose energy is correlated with the angle-dependent radial flow. 
We compare this flow for $pp$, $p$Pb, $p$O, OO and PbPb collision systems and conclude in particular that the $p$Pb flow is large and anisotropic. 
The collective medium pushes out the hard partons, and more so in $p$Pb, or in-plane, than in $pp$ collisions, or out-of-plane. 
The mechanism qualitatively agrees with experiment and predicts a significantly smaller elliptic flow signal for jets than for hadrons.
\end{abstract}

\maketitle

{\bf Introduction - } While the Higgs particle is undoubtedly the biggest discovery of the Large Hadron Collider (LHC) \cite{CMS:2012qbp, ATLAS:2012yve}, the most surprising discovery is likely the observation of collectivity even in high-multiplicity $pp$ collisions \cite{CMS:2016fnw}. 
``Collectivity'' here means multiple interactions between partons beyond the perturbative few-body interactions and was identified by non-trivial correlations of two particles separated my many units of pseudorapidity\cite{Grosse-Oetringhaus:2024bwr}.
This observation is furthermore seen in $p$Pb collisions and even underpins the enhancement of strangeness as a function of multiplicity \cite{ALICE:2016fzo}. 

This Letter addresses a long-standing conundrum: at high transverse momentum ($p_T$), hadrons exhibit azimuthal anisotropy ($v_2$), and are hence correlated with the event plane.
In large AA systems, the traditional explanation involves anisotropic energy loss of partons traversing the quark-gluon plasma (QGP).
Puzzlingly, in small systems such a $v_2$ is also seen, even though the nuclear modification factor $R_{p\text{A}}$ is consistent with unity and hence does not indicate any energy loss \cite{CMS:2016xef,ATLAS:2016xpn,ALICE:2018vuu,ALICE:2021est}.
This is perhaps unsurprising, as even in large systems it is a famous and longstanding problem \cite{Noronha-Hostler:2016eow,Andres:2019eus} to quantitatively describe $R_\text{AA}$ and $v_2$ simultaneously. 
The key insight there is that early-time energy-loss is isotropic.
Not only have these puzzles occupied the community for over a decade, but it ultimately became one of the main motivations for the current light-ion at the LHC \cite{Brewer:2021kiv}.

Here we propose a qualitatively new mechanism that helps for the puzzle in PbPb and resolves the high-$p_T$ $p$Pb regime. 
Any observed hadron must consist of at least two partons of which at least one must come from the hydrodynamic or collective medium. 
Advanced hydrodynamic simulations tell us that the radial flow of such a medium is significantly larger in PbPb or $p$Pb collisions than for $pp$ collisions. The radial flow is moreover anisotropic in both the PbPb and $p$Pb cases.
We will show that, due to the anisotropic energy \textit{gain} by the radial boost, it is natural that the $R_{p\text{A}}$ is around unity and that the $v_2$ can be significant even up to a $p_T$ of $50\,$GeV.

While the main motivation for the present study was the $R_{p\text{A}}$-$v_2$-puzzle in $p$Pb and PbPb, our resolution of the problem gains more mileage.
The coalescence with radial flow reproduces the newly measured $R_\text{OO}/R_{p\text{O}}^2$ ratio\cite{ALICE:2026zck}, which is constructed to be virtually independent of nPDF effects and hence theoretically an especially clean observable \cite{Jonas:2026yoz}. 

{\bf Radial flow - } 
We start with the radial flow velocities as a function of angle with respect to the event plane (the event plane defined by the angle of $Q_2 = \sum e^{2i\phi_i}$, with $\phi_i$ the angles of all hydrodynamic particles produced at mid-rapidity). Since we are interested in its effect on high-$p_T$ hadrons at hadronization we evaluate this radial fluid velocity on the freeze-out surface as the the velocity in the direction of a path that started at some binary collision location in an arbitrary direction \cite{Beattie:2022ojg}.

\begin{figure}[t]
        \centering
        \includegraphics[width=0.9\linewidth]{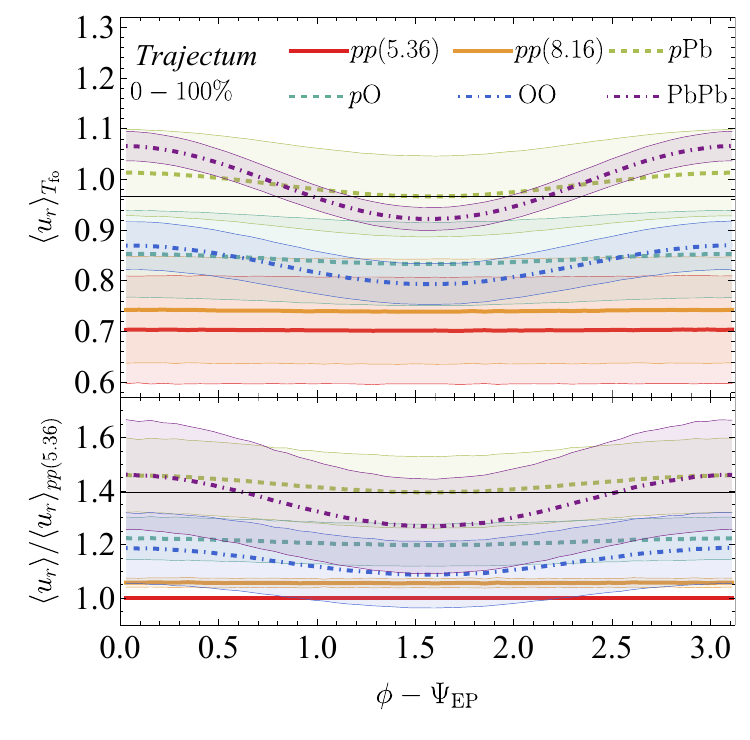}
        \caption{The radial velocity (top) and the ratio with $pp$ (bottom) in the direction of the outgoing parton on the freeze-out surface as a function of angle for $pp$, $pp$, $p$Pb, $p$O, OO and PbPb at $\sqrt{s_\text{NN}}$ of $5.36,\,8.16,\,8.16,\,9.12,\,5.36$ and $5.02\,$TeV respectively. Especially the $p$Pb radial flow is enhanced and anisotropic, also when comparing to the $pp$ at $8.16\,$TeV. To guide the eye, a reference line is shown at the $p$Pb minimum mean value. }
        \label{fig:uradial}
    \end{figure}

Fig.~\ref{fig:uradial} shows these radial velocities $u_r$ for six colliding systems at minimum bias coming from the \emph{Trajectum} framework \cite{Nijs:2020ors,Nijs:2020roc}. Especially for small systems such as $pp$, $p$O or $p$Pb there is considerable systematic uncertainty on this radial velocity profile. It depends sensitively on the initialization time $\tau_0$, the nucleon size, sub-nucleonic structure, viscosities, freeze-out temperature $T_\text{fo}$, and other parameters. Within the \emph{Trajectum} framework, these uncertainties are included by evaluating all our results on ten likely parameter settings (from a Bayesian analysis) that consistently vary all these parameters \cite{Nijs:2021clz}. Importantly, we thus evaluate all our computations ten times, but often find that for final observables, such as $R_{p\text{A}}$, much of the uncertainty cancels when taking ratios. %

In this work we included one extra uncertainty: the starting point of the path at the binary collision location. In previous work this was always the average of the positions of the two colliding participant nucleons. To be more realistic we now introduced a Gaussian smearing of this location of width $0,$ $0.25\,$ and $0.5\,$fm and likewise treated those three options as a systematic uncertainty. This is the dominant uncertainty for $pp$ collisions in Fig.~\ref{fig:uradial}. 

The most important insight from Fig.~\ref{fig:uradial} is that the $p$Pb radial velocity is much larger than in $pp$ collisions and that it is moreover about 5\% larger for in-plane than for  out-of-plane paths. Both of these observations can be understood as coming from a tiny, but explosive droplet of anisotropic QGP formed in the $p$Pb collision. 

{\bf Model - } 
Our model assumes a hard parton losing energy when traversing a QGP and then at freeze-out combining with a thermal parton at that point. Motivated by elastic energy loss scaling as degrees of freedom times $T^2$ we chose $s/T$ in the fluid restframe as integrand for the local energy lost\footnote{We verified that almost indistinguishable results are obtained chosing $T^3$ instead of $s/T$.}, with $s$ the entropy density and $T$ the temperature (see also \cite{vanderSchee:2023uii,vanderSchee:2025hoe,Stojku:2020tuk}). The proportionality factor $\kappa(p_T)$ is ultimately fixed by a reference dataset in the 0-5\% centrality class of the charged hadron or pion PbPb nuclear modification factor \cite{CMS:2016xef, vanderSchee:2025hoe, ALICE:2019hno}. 
The energy of the thermal parton is given by a boosted Boltzmann value $e_\text{th}[u_r]$ given below. Crucially, we assume the same model for all systems involved, from $pp$ to PbPb. 

The nuclear modification factor $R_{AA}(p_T)$ is then determined by an integral over paths $\mathcal{I}$ starting from the same binary collision location and ending at the freeze-out surface as for $u_r$:
    \begin{align}
    R_{AA}(p_T) & = \frac{\int_\text{AA}\mathcal{D}\mathcal{I}\mathcal{D}u_r\,\sigma(p_{T}+\kappa[p_{T}]\, \mathcal{I}-e_\text{th}[u_r])}
    {\int_{pp}\mathcal{D}\mathcal{I}\mathcal{D}u_r\sigma(p_{T}+\kappa[p_{T}]\,\, \mathcal{I}-e_\text{th}[u_r])},\label{eq:path-length}\\
    \mathcal{I} &\equiv \int_{\tau_0}^{\tau_{fo}} s/T\, u_\mu dL^\mu,\quad e_\text{th}[u_r] = \frac{3T_{fo}}{\sqrt{1+u_r^2}-u_r}, \nonumber 
    \end{align}
where $\sigma(p_T)$ is approximated by the charged hadron spectrum in $pp$ collisions \cite{CMS:2016xef} (the results are not really sensitive to $\sigma(p_T)$ since the same spectrum enters numerator and denominator \cite{vanderSchee:2025hoe}). The current study takes into account fluctuations in $\mathcal{I}$ (unlike \cite{vanderSchee:2025hoe}) and $u_r$, but not a possible correlation between them. Paths that start below $T_\text{fo}$ are assigned $u_r = 0$.

    \begin{figure}[t]
        \centering
        \includegraphics[width=0.9\linewidth]{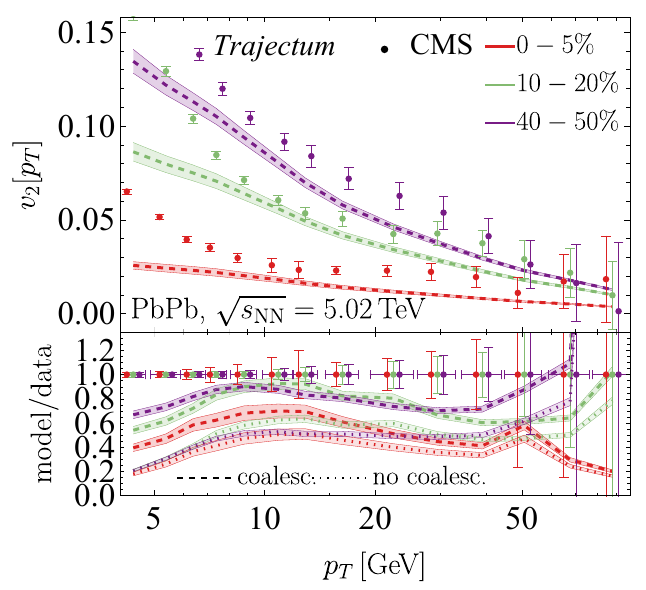}
        \caption{Traditionally, high-$p_T$ $v_2$ has been interpreted as pathlength dependent energy loss. Here we show that this is not sufficient and that coalescence with a radial flow profile as in Fig.~\ref{fig:uradial} can account for half of the $v_2$. Energy loss plus coalescence is mostly consistent with CMS data \cite{CMS:2017xgk} from about $p_T=8\,$GeV from semi-central to peripheral events.}
        \label{fig:PbPbv2}
    \end{figure}

In practice, for example for $p$Pb, we have to evaluate Eqn.~\ref{eq:path-length} consistently for PbPb (to fit the 0-5\% reference) as well as $p$Pb and $pp$ (we take the closest energy available) and all of this ten times to estimate the systematic uncertainty. We comoute the azimuthal anisotropy $v_2$ using $v_2 = \frac{\pi}{4}\frac{R_\text{in}-R_\text{out}}{R_\text{in}+R_\text{out}}$ with $R_\text{in/out}$ the corresponding $R_\text{AA}$s evaluated for in- and out-of-plane paths and radial velocities.

In this work we do not take into account nPDFs, but note that they are not expected to significantly affect the $v_2$
and even the nPDF effects on $R_{p\text{O}}$ and $R_\text{OO}$ are consistent with unity within uncertainties for $p_T$ above approximately $8\,$GeV \cite{Jonas:2026yoz}. 
When fitting ALICE $\pi_0$ measurements we take instead the 0-5\% charged pion $R_\text{AA}$ \cite{ALICE:2019hno} as reference to fit $\kappa(p_T)$.

The integrals computed in Eqn.~\ref{eq:path-length} were relatively straightforward extensions of existing \emph{Trajectum} technology \cite{Nijs:2020ors,Nijs:2020roc,Beattie:2022ojg} and were implemented with help of Claude Opus 5.

{\bf Flow and coalescence from small to large - }
Perhaps surprisingly, including coalescence does not change the PbPb $R_\text{AA}$ as presented in \cite{companion} and reproduced in the appendix. Most of this is explained by the refitting of $\kappa(p_T)$ in the 0-5\% centrality class. 
However, for the azimuthal ansitropy $v_2$ the effect is dramatic. Including coalescence is mostly consistent with experiment for centralities $10\%$ to $50\%$ and for $p_T$ above about $8\,$GeV, whereas only including energy loss would undershoot experiment by more than a factor of two (see \cite{companion} and dotted curves in ratio of Fig.~\ref{fig:PbPbv2}).

\begin{figure}[t]
        \centering
        \includegraphics[width=0.9\linewidth]{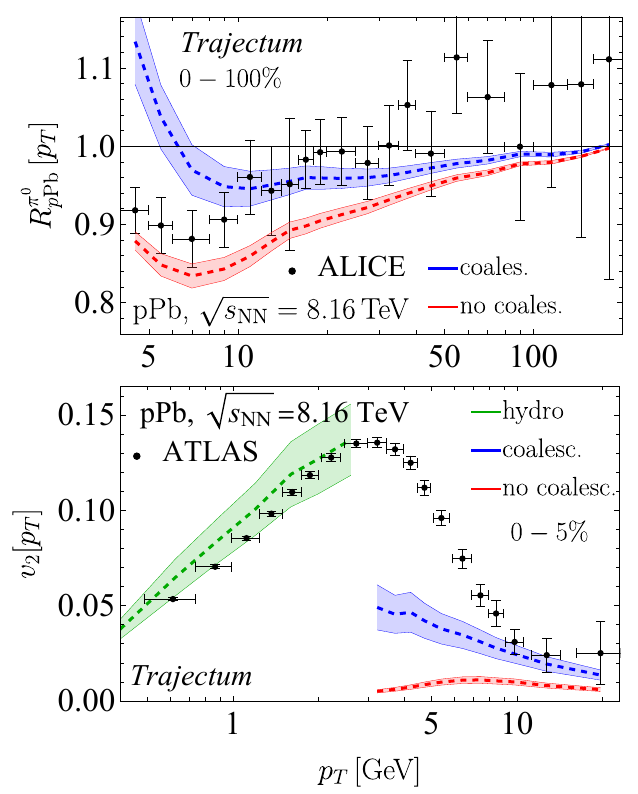}
        \caption{(top) We show the $R_{p\text{Pb}}^{\pi^0}$ with and without coalescence as compared with ALICE data \cite{ALICE:2021est}. The larger $p$Pb radial flow velocity gives a boost to the $\pi_0$ in such a way that it approximately compensates the energy loss present in the no coalescence case. (bottom) In green we show the standard soft $p_T$ elliptic flow from the same hydro simulations. Since energy loss is small and approximately isotropic the no coalescence curve (red) gives an almost negligible $v_2$ (note the different scale in $p_T$). The radial flow is anisotropic (see Fig.~\ref{fig:uradial}) and due to the steep spectrum even a tiny in-plane boost gives rise to a significant $v_2$ all the way up to $20\,$GeV (blue) that compares well with ATLAS \cite{ATLAS:2019vcm}. It is remarkable that above $10\,$GeV both $R_{p\text{Pb}}^{\pi^0}$ and $v_2$ simultaneously describe the data. }
        \label{fig:RAAv2pPb}
    \end{figure}

The most profound impact is on $p$Pb $R_{p\text{Pb}}$ and $v_2$. As shown in Fig.~\ref{fig:RAAv2pPb}, including radial flow and coalescence allows for a simultaneous description above $p_T=9\,$GeV and %
provides a mechanism to have a $R_{p\text{Pb}}$ around unity while at the same time allowing a significant $v_2$. 

Lastly, recently ALICE measured for $\pi_0$ the $R_{p\text{O}}$, $R_\text{OO}$ and the ratio $R_\text{OO}/R_{p\text{O}}^2$. Since our current model does not include nPDF effects the ratio is particularly important since it is constructed in such a way not to depend on nPDFs.
As shown in \cite{companion} and reproduced in the appendix it is difficult to describe the ratio using only energy loss\footnote{Both hybrid \cite{Kudinoor:2026wcs} and the Cape Town model \cite{Bert:2026uxa} do quite well in \cite{ALICE:2026zck}; as far as we understand, however, they assume  $R_{p\text{O}} = 1$, whereas in our model this follows within self-consistent model assumptions among all colliding systems.}. Coalescence significantly boosts the $R_{p\text{O}}$ and makes the ratio consistent with ALICE all the way down to $p_T=4\,$GeV. It is tempting to speculate that the p$O$ and $OO$ deviations below $8\,$GeV  are due to nPDF effects as also suggested by \cite{Gebhard:2024flv,Mazeliauskas:2025clt,Jonas:2026yoz}.

    \begin{figure}[t]
        \centering
        \includegraphics[width=0.9\linewidth]{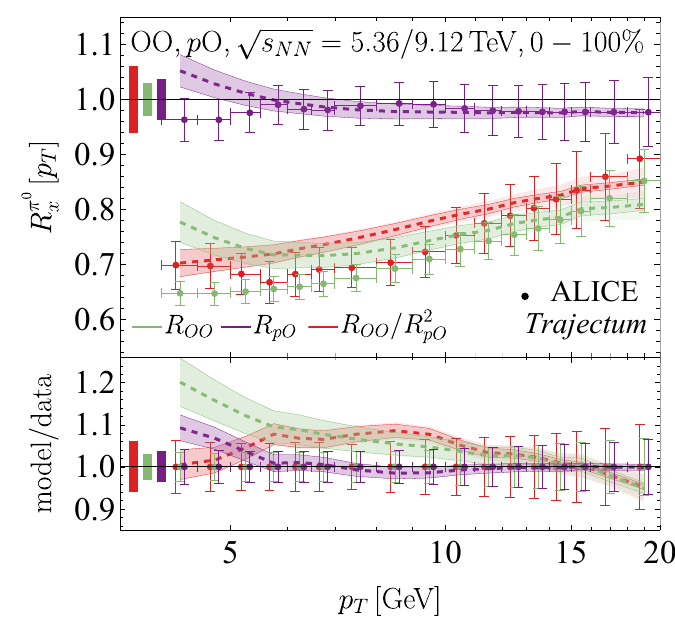}
        \caption{Consistent with Fig.~\ref{fig:RAAv2pPb} we find that coalescence dramatically improves the agreement with ALICE data \cite{ALICE:2026zck} for all three $R_{p\text{O}}$, $R_\text{OO}$ and $R_\text{OO}/R_{p\text{O}}^2$, which was challenging to describe with only energy loss \cite{companion}. Note that this work includes neither nPDF effects nor uncertainties, but that $R_\text{OO}/R_{p\text{O}}^2$ is constructed such that these are negligible \cite{Jonas:2026yoz}.}
        \label{fig:ROOopO2}
    \end{figure}

{\bf Discussion - }
From the moment $R_{p\text{Pb}}$ was measured to exceed unity it caused confusion \cite{ALICE:2012mj}. Even if a QGP produced in $p$Pb was tiny, surely an energetic parton could not gain energy? With nPDF effects known to be relatively small, the $R_{p\text{Pb}}$ was almost impossible to reconcile with the positive elliptic flow. %

If one is poetic, one can now understand high-$p_T$ $p$Pb in a new way. Indeed, the parton could lose a bit of energy. But a parton is never alone: due to confinement, it has to find a parasitic partner, and this partner can, in fact, gain energy by flowing inside an accelerating expanding plasma. If one takes the hydrodynamic paradigm in $p$Pb collisions seriously, it is then also natural that this energy gain is anisotropic, in reflection of the anisotropic radial velocity presented in Fig.~\ref{fig:uradial}. Quantitatively, the energy gain is small ($50-200\,$MeV), but due to the steeply falling spectrum it is still important even at a $p_T$ of over $20\,$GeV.
Perhaps, in these times, it is not surprising that most solutions focused on the energetic high-$p_T$ parton and neglected the typical but fast-moving partner from the collective.

Traditionally, coalescence has been interpreted as the driving force in the $p_T=3-8\,$GeV region.  At higher $p_T$, partons would fragment into hadrons \cite{Molnar:2003ff}. In the Lund string hadronization model such fragmentation would still be colour connected to the other particles, or even the beam. In our model the colour connection is to a thermal parton. We stress, however, that all three pictures could be relevant. The only requirement for our model to have explanatory value is that hadronization feels the collective motion around the parton\footnote{A sensible alternative to our thermal parton would for instance be to take a comoving thermal parton (similar to \cite{Zhao:2021vmu} or to heavy flavour coalescence \cite{Zhao:2023nrz}). This would presumably increase the effect here presented.}. 

\begin{figure}
    \centering
    \includegraphics[width=0.9\linewidth]{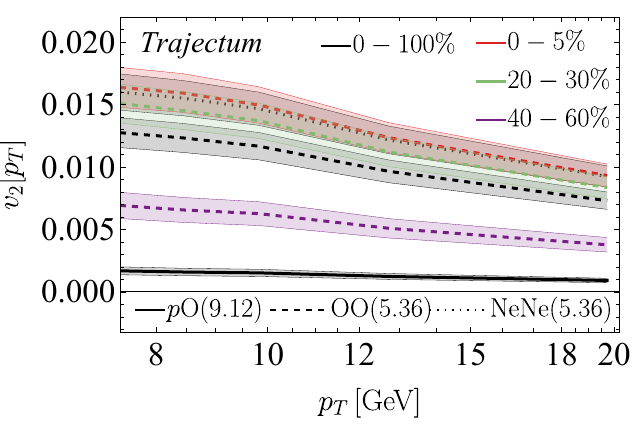}
    \caption{We apply the same model for $p$O, OO and NeNe collisions to predict the $v_2$ for minimum bias and centrality selected (OO). The NeNe curve overlaps almost identically with OO at $0-5\%$  due to the bowling pin shape of $^{20}$Ne \cite{Giacalone:2024luz}.}
    \label{fig:v2pOandOO}
\end{figure}

Various attempts have been made to simultaneously describe high-$p_T$ $R_{\text{PbPb}}$, $R_{p\text{Pb}}$ and $v_2$ \cite{Xu:2015bbz,Noronha-Hostler:2016eow,Zigic:2019sth,Andres:2019eus,Zhao:2021vmu,Pablos:2025cli,Soudi:2023epi,Datta:2025gql}, even including coalescence \cite{Zhao:2021vmu}. There, however, coalescence was limited to below $10\,$GeV and even there effects were more moderate than here. Coalescence also was not connected to an enhanced $R_{p\text{Pb}}$. 
The works \cite{Kolbe:2015rvk,Faraday:2023mmx,Faraday:2024gzx,Berg:2026tnk} suggest energy gain as a mechanism for enhancing $R_{p\text{Pb}}$, but without involving collective radial motion or coalescence.
Our implementation provides a higher $v_2$ without invoking initial state correlations, nPDF effects, or subtle experimental soft-hard correlations. The model has predictive power, such as for the $v_2$ for $p$O, NeNe and OO for $p_T > 7\,$GeV (shown in Fig.~\ref{fig:v2pOandOO}).

The weighted path model with simplified coalescence should be improved with a full parton shower and a more realistic hadronization. It would also be worthwhile to investigate alternatives to compute $u_r$ that are not hydrodynamic such as partcon cascades like AMPT \cite{Lin:2004en}.
But the scarcity of real solutions to the high-$p_T$ $v_2$ puzzles suggests that radial collective motion will play an essential role in any model describing high-$p_T$ hadrons in small systems. Conversely, jets are famously less sensitive to hadronization. As such we can predict that both the jet $R_\text{AA}$ and $v_2$ in such small systems will be smaller than the corresponding hadron ones. Indications for this can be seen in peripheral PbPb collisions \cite{Pablos:2025cli,companion}.

{\bf Acknowledgments - }
    The authors are grateful for useful discussions with Coleridge Faraday, Florian Jonas, Aleksas Mazeliauskas, Govert Nijs, Rosi Reed and Urs Wiedemann.
    This work is supported by the DST/NRF in South Africa under Thuthuka grant number TTK240313208902.

\bibliography{coalescence,manual}

\clearpage

\clearpage

\onecolumngrid

\section{Appendix}

    \begin{figure}
        \centering
  \begin{minipage}[c]{0.32\textwidth}
    \includegraphics[width=\linewidth]{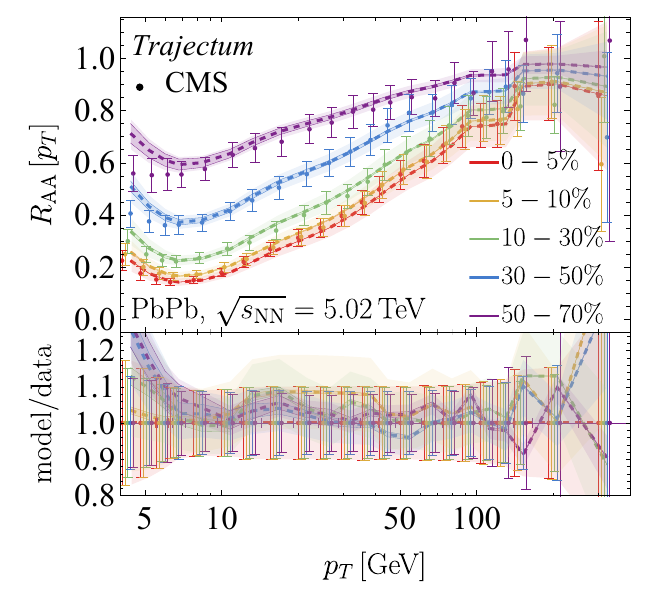}
  \end{minipage}\hfill
  \begin{minipage}[c]{0.32\textwidth}
    \includegraphics[width=\linewidth]{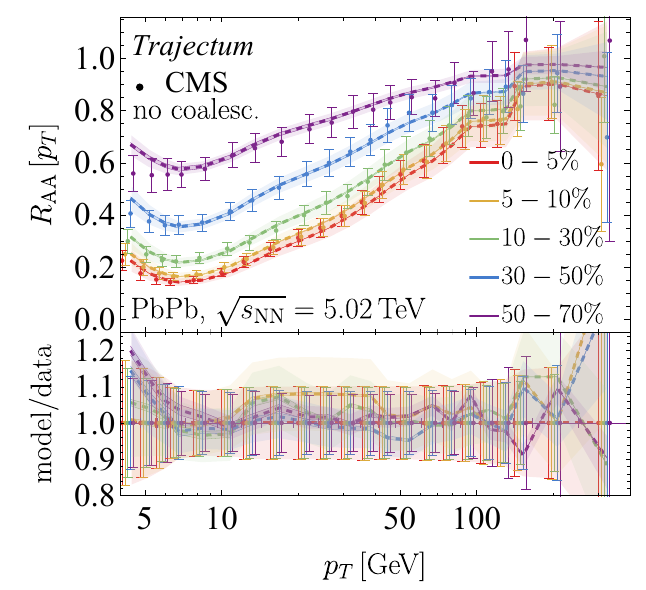}
  \end{minipage}\hfill
  \begin{minipage}[c]{0.32\textwidth}
    \includegraphics[width=\linewidth]{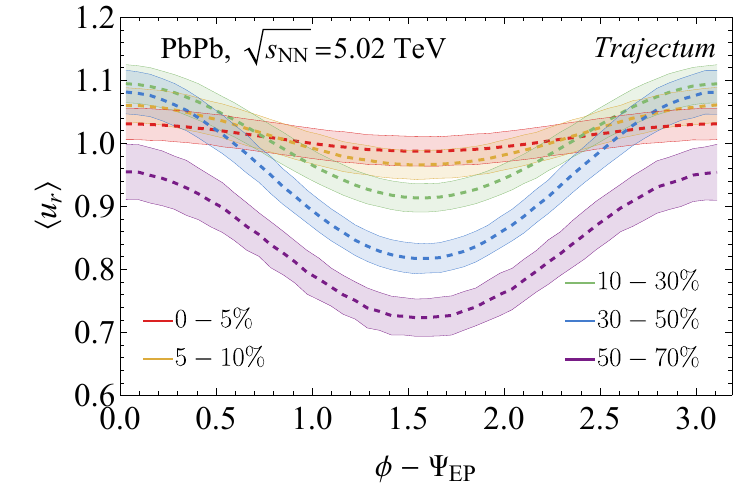}
  \end{minipage}
        \caption{We show the PbPb $R_\text{AA}$ for several centrality classes as compared with CMS data \cite{CMS:2016xef} with (left) and without (middle) the coalescence as described in the main text. Note that by construction the 0 - 5\% centrality class matches the data by fitting the coefficient $\kappa(p_T)$. (right) We show the PbPb radial velocity as in Fig.~\ref{fig:uradial}, but now for the same centrality classes as in the two plots on the left.}
        \label{fig:PbPbRAA}
    \end{figure}

For completeness we reproduce here the PbPb $R_\text{AA}$ of \cite{companion}, both without (as in the reference) and including coalescence. Even though for this $R_\text{AA}$ the differences are hardly visible the main text shows that coalescence is essential to explain the subtle differences in the in- and out-of-plane $R_\text{AA}$ that are ultimately responsible for the high-$p_T$ azimuthal anisotropy.

Perhaps it is a bit surprising that coalescence makes such little difference in the PbPb $R_\text{AA}$. Mostly this is due to the fact that we fit the 0-5\% centrality class, but that does not fully explain the most peripheral class.

For completeness we show the equivalent of Fig.~\ref{fig:ROOopO2} in the main text but with coalescence turned off (see also \cite{companion}).

    \begin{figure}[h]
        \centering
        \includegraphics[width=0.4\linewidth]{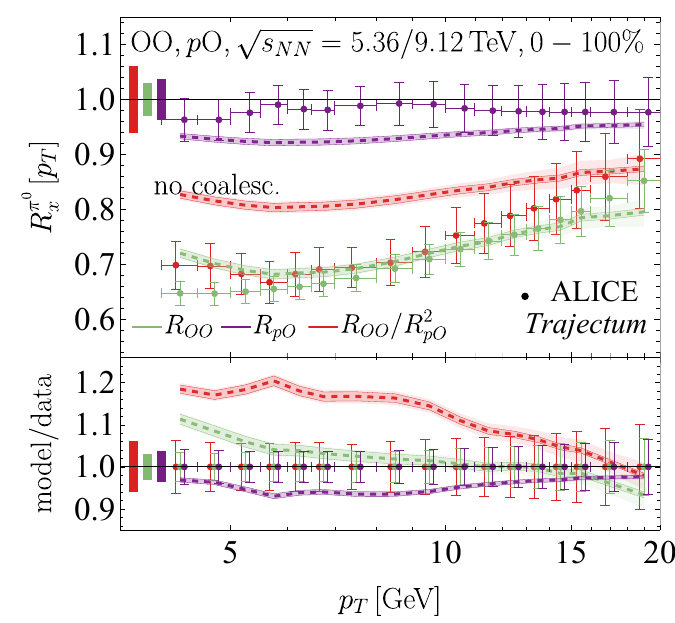}
        \caption{For completeness we reproduce the no coalescence figure from \cite{companion}. This has to be compared with Fig.~\ref{fig:ROOopO2} in the main text.}
        \label{fig:ROOopO2nocoal}
    \end{figure}

\end{document}